\documentclass[prl,twocolumn,showpacs,showkeys,preprintnumbers,amsmath,amssymb]{revtex4}

\usepackage{graphicx}
\usepackage{graphics}
\usepackage{dcolumn}
\usepackage{bm}
\usepackage{amssymb}
\usepackage{amsmath}
\begin{document}
\title{Very High Mach Number Electrostatic Shocks in Collisionless Plasmas}
\author{G. Sorasio}
\email{gsorasio@ist.utl.pt}
\altaffiliation[Also at ]{ISCAT s.r.l., V. S. Pellico 19, 12037 Saluzzo, Italy.}
\author{M. Marti}
\author{R. Fonseca}
\altaffiliation[Also at ]{DCTI/ISCTE, 1649-026 Lisbon, Portugal.}
\author{L. O. Silva}\
\email{luis.silva@ist.utl.pt}
\affiliation{GoLP/ Centro de F\'{\i}sica dos Plasmas, Instituto Superior T\'ecnico,\\ Av. Rovisco Pais, 1049-001 Lisbon, Portugal.}
\date{\today}

\begin{abstract}
The kinetic theory of collisionless electrostatic shocks resulting from the collision of plasma slabs with different temperatures and densities is presented. The theoretical results are confirmed by self-consistent particle-in-cell simulations, revealing the formation and stable propagation of electrostatic shocks with very high Mach numbers ($M \gg 10$), well above the predictions of the classical theories for electrostatic shocks.
\end{abstract}

\maketitle
The collision of clouds of plasma with different properties (temperature, density, composition etc.) is a  scenario quite common in nature. For instance, during supernovae explosions, large quantities (10 solar masses) of high temperature plasma ( $T \sim  10^{6-8} \, \mathrm{K}$) are ejected into the interstellar medium ($n \sim 1 \, \mathrm{cm}^{-3}$, $T \sim 10^{2-4} \, \mathrm{K}$)  \cite{Longair,Gaisser}, and plasma cloud collisions are at the core of the fireball model for gamma ray bursters \cite{Piran}. Plasma cloud collisions also occur when the solar wind interacts with the Earth Magnetosphere, or when it encounters the interstellar medium in  the heliosphere region \cite{Krimigis}. In the laboratory, such scenarios appear during the laser induced compression of plasma foils in solid targets \cite{Silva}. 

The collision of plasma shells leads to the onset of plasma instabilities and to the development of nonlinear structures, such as solitons, shocks and double layers \cite{Schamel}. In the absence of an ambient magnetic field, the shock waves are electrostatic \cite{Fors1,Fors2}, and the dissipation is provided by the population of electrons trapped beyond the shock front \cite{Schamel,MongJoyce} and, for stronger shocks, by the ion reflection from the shock front \cite{Sagdeev}. Whilst the properties of shocks induced by collision of identical plasma shells, or by compression of plasma clouds, have been extensively studied in the past \cite{Moiseev,Schamel,Sagdeev,MongJoyce,Fors1,Fors2,Bisk,BiskPark,Bardotti,Silva}, the properties of the electrostatic shock waves formed during the collision of diverse plasma slabs of arbitrary temperature and density are rather unexplored \cite{Ishiguro}.        
The theory for electrostatic shocks induced by impact of identical plasma shells predicts an absolute maximum Mach number $M_\mathrm{max} \simeq 3$ (or, when ion reflection and thermal effects are included, $M_\mathrm{max}^* \simeq 6$). However, collisionless shock waves with Mach numbers ranging between $10$ and $10^3$ have been observed in many astrophysical scenarios \cite{Gaisser}, and very large Mach number cosmic shock waves are thought to play a crucial role in the evolution of the large scale structure of the Universe \cite{Miniati,Ryu}.    

In this Letter, we present a kinetic theory describing the properties of the very high Mach number  ($M\gg10$) laminar shock waves arising from the collision of slabs of plasma with different properties (temperatures, densities), and in the absence of an ambient magnetic field. 
We demonstrate that the shock properties are strongly influenced by $\Theta$, the ratio of the electron temperatures in the two slabs, and by $ \Upsilon$, the ratio of the electron densities in the two slabs. The analysis shows that when the electron temperature $T_\mathrm{e\,R}$ of the downstream slab (R) is higher than the electron temperature  $T_\mathrm{e\,L}$ of the upstream slab (L),
the shock waves can have very large Mach numbers, which are otherwise not supported by isothermal plasmas \cite{Fors1,Fors2, Schamel}. 
The model predicts that the maximum allowed Mach number increases with $\Theta$, without an absolute upper limit. The theoretical results are confirmed by one-dimensional (1D) self-consistent particle-in-cell simulations, demonstrating the formation and the stable propagation of electrostatic shock waves with very large Mach numbers ($\mathrm{M} ~\sim 20$). 

\begin{figure}
\includegraphics[width=2.5in]{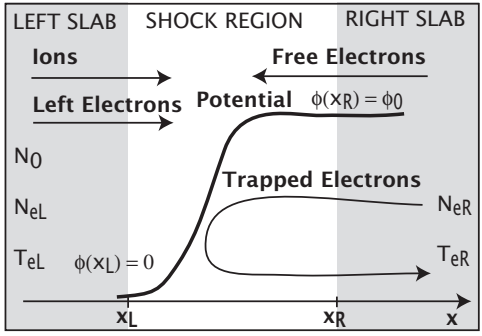}
\caption{\label{fig:cartoon} Geometry of the collisionless laminar shock wave. The bold line represents the electrostatic potential. The electrons from the slab L move freely, while the electrons from the slab R can be either free or trapped. The ions flow towards the shock, and are decelerated by the potential. }
\end{figure}

The shock transition region is modeled as a planar one dimensional double layer, which is stationary in the reference frame of the shock; the electrostatic potential increases monotonically from $\phi=0$ at $x = x_\mathrm{L}$ to $\phi=\phi_0$ at $x = x_\mathrm{R}$, as shown in Figure \ref{fig:cartoon}. The one dimensional treatment should hold for propagation distances smaller than  the transverse dimension of the shock.  The structure of the double layer is maintained by two populations of free electrons (from the L and R slabs), a population of trapped electrons (from the R slab), and a population of cold ions. The model considers the contribution of the trapped particles self-consistently, by treating the electrons in a kinetic fashion \cite{Schamel,Fors1}. To maintain a steady state, the number of electrons in unit time $dt$, with velocity between $[v_e,v_e+dv_e]$ and position between $[x,x+dx]$, must balance the electrons injected from the left slab ($L$), at $x=x_\mathrm{L}$, with velocity between $[v_L,v_L+dv_L]$, and the electrons from the right slab ($R$), at $x=x_\mathrm{R}$, with velocity between $[v_R,v_R+dv_R]$. We can express the electron distribution function $f_e(x,v_e)$, at any given point $x$, as a function of the electrons injected at the left and right boundaries, which follow known distribution functions. The electron velocity, which results from conservation of energy, can be written as $v_e=\sqrt{v_L^2+\frac{2 e \phi}{m_e}}=-\sqrt{v_R^2+\frac{2 e (\phi-\phi_0)}{m_e}}$, depending if the particles are arriving from the left boundary or from the right boundary; here $m_e$ is the electron mass, and $e$ is the elementary charge. Assuming  that only electrons with positive velocity enter from the  slab L, and only electrons with negative velocity enter from the slab R, we obtain:
\begin{equation}
f_e(x,v_e)\,dv_e=\left. \frac{v_L}{v_e} f_\mathrm{L}(v_L) \,dv_L \right|_0^\infty +\left. \frac{v_R}{v_e} f_\mathrm{R}(v_R) \,dv_R\right|_{-\infty}^{0} \label{eq:fe},
\end{equation}
where  $f_\mathrm{L}(v_L)$ and $f_\mathrm{R}(v_R)$ are the electron distribution function in the L and R slabs.  
We assume that the electrons in the L slab follow the drifting Maxwell-Boltzmann (MB)  distribution function \cite{Schamel}
$ f_\mathrm{L}(v_L)=\frac{N_{0L}}{V_{TeL} \sqrt{2 \pi}} e^{-(v_L-V_i)^2 / V_{TeL}^2}$, and we consider that the electron thermal velocity $V_{Te\alpha}=\sqrt{k_B T_\mathrm{e\,\alpha}/m_e}$ is greater than the shock speed $V_i$.  Here $k_\mathrm{B}$ is the Boltzmann constant, while $T_\mathrm{e\,\alpha}$ and $N_{0\alpha}$ are the  electron temperature and density in the slab $\alpha$. In the R slab, the electrons are affected by the potential $\phi_0$, and their distribution function is composed of a free and a trapped part, $f_\mathrm{R}(v_R)=f_\mathrm{Rf}(v_R)+f_\mathrm{Rt}(v_R)$. If their kinetic energy is larger than the electrostatic energy (i.e. $|v_R|>|v_c|=|\frac{2 e \phi_0}{m_e}|$), they are free, and continuously decelerate while moving towards the left boundary. Following \cite{Fors1,Schamel}, we assume that the free electrons can be modeled by the distribution function $ f_\mathrm{Rf}(v_R)=\frac{N_{0R}}{V_{TeR} \sqrt{2 \pi}} e^{-\frac{v_R^2}{V_{TeR}^2}+\frac{e \phi_0}{k_B T_\mathrm{eR}}}$. The electrons with kinetic energy  lower than the electrostatic energy (i.e. $|v_R|<|v_c|$) are trapped, and are assumed to follow the flat top distribution function  $f_\mathrm{Rt}(v_R)=\frac{N_{0R}}{V_{TeR} \sqrt{2 \pi}}$. This idea of describing  an electron gas, composed of free and trapped particles, by using a  MB distribution function with a  flat-top, has been widely used in the past to model stationary ion acoustic structures  \cite{Schamel,Fors1,Fors2,Bisk}, and it is called `maximum-density-trapping' approximation \cite{MongJoyce,Schamel}. 

The density of electrons along the shock can be calculated by integrating the electron distribution function $f_e(x,v_e)$ in velocity space. Following Equation (\ref{eq:fe}), we can separate the electron density along the shock in the contributions from the L and R slabs. 
The density of the electrons coming from the slab L can be written as  $n_\mathrm{L}(\varphi)= \frac{N_{0L}}{2}e^{\varphi}\mathrm{Erfc}\sqrt{\varphi}$,
where $\varphi=\frac{e \phi}{k_B T_\mathrm{eL}}$. 
In our model, the electrons coming from the region L are continuously accelerated while moving towards the right boundary, and are not reflected or trapped by the electrostatic potential.    
The density of electrons from the slab R is obtained using Equation (\ref{eq:fe}),  integrating the term containing the free electron distribution, $f_\mathrm{Rf}(v_R)$, in the range $[-\infty,-v_c]$, and the term containing the trapped electron distribution, $f_\mathrm{Rt}(v_R)$, in the range $[-v_c,0]$, leading to 
$n_\mathrm{R}(\varphi)= \frac{N_{0L} \Upsilon}{2}e^{\varphi / \Theta}\mathrm{Erfc}\sqrt{\varphi / \Theta} + \frac{2 N_{0L} \Upsilon}{\sqrt{\pi}}\sqrt{\varphi/\Theta}$, where $\Upsilon=N_\mathrm{0R}/N_\mathrm{0L}$ is the density ratio.

In the present model, the ions are cold, flowing towards the shock with  velocity $V_i$, and being continuously decelerated by the electrostatic potential. The ion density is determined by considering the energy and the mass conservation equations \cite{BiskPark}, and can be written as $n_\mathrm{i}=N_0/\left( \sqrt{1-2 \varphi/M^2}\right)$, where $M=V_i/V_s$ is the ion acoustic Mach number, $V_s=\sqrt{k_B T_\mathrm{eL}/m_i}$ is the ion sound speed, $m_i$ is the ion mass, and $N_0$ is the unperturbed ion density in the slab L. The reflection of ions is not included in the present model. Such assumption is consistent with a double layer solution maintained by a population of trapped electrons \cite{MongJoyce,Schamel}. The aim of the present Letter is to show that the collision of plasma slabs, with appropriate temperature and density ratios, leads to shock waves with very large Mach numbers. Since the occurrence of ion reflection increases even further the velocity of the shocks, the conclusions of the present Letter are to be considered as conservatives. 

In the derivation of the ion and electron densities, we used the quantities $N_\mathrm{0R}$, $N_\mathrm{0L}$ and $N_\mathrm{0}$, which can now be evaluated by applying boundary conditions proper of double layers \cite{MongJoyce,Schamel}. Using charge neutrality at $x=x_\mathrm{L}$ and at $x=x_\mathrm{R}$, we obtain the conditions $N_0=N_\mathrm{0L}+N_\mathrm{0R}$ and $n_\mathrm{R}(\varphi_0)+n_\mathrm{L}(\varphi_0)=n_\mathrm{i}(\varphi_0)$, thus leading to $N_\mathrm{0L}$ and $N_\mathrm{0R}$ as  function of the unperturbed ion density $N_0$, of the normalized potential $\varphi_0$, and of the Mach number $M$. Since it is clear that the electron distribution functions are always positive, we must apply the physical inequalities $N_\mathrm{0L},N_\mathrm{0R}>0$.    

By combining the ion and electron densities with Poisson's equation, and since the dynamics of the electrostatic potential is analogous to the motion of a particle in a  potential well $\Psi$, we find that  the evolution of the electrostatic potential is governed by $\frac{1}{2}\left( \frac{\partial \varphi}{\partial \chi}\right)^2+\Psi(\varphi)=0$ \cite{Tid} , where the spatial coordinate, $\chi$, is normalized to the electron Debye length $\lambda_d=\sqrt{k_B T_\mathrm{eL} / 4 \pi e^2 N_0}$, and the nonlinear Sagdeev  potential \cite{Sagdeev} is:
\begin{equation}
\Psi(\varphi)=  -\left\{P_e(\varphi,\Theta,\Upsilon)-P_\mathrm{I}(\varphi, M)\right\}. \label{eq:pot}
\end{equation}
Here $P_e(\varphi,\Theta,\Upsilon)=P(\varphi,\Theta,\Upsilon)-P(\varphi=0,\Theta,\Upsilon)$ is the electron pressure, and $P_\mathrm{I}(\varphi,M)=M^2\left(1- \sqrt{1-2\varphi/M^2}\right)$ is the  ion pressure, normalized to $N_0 k_B T_\mathrm{eL}$. The pressure term $P(\varphi,\Theta,\Upsilon)=P_\mathrm{L}+P_\mathrm{R}$ includes the contribution  of the electrons from the slab L, 
$P_\mathrm{L}(\varphi)= \frac{N_\mathrm{0L}}{2 N_0}\left[e^{\varphi}\mathrm{Erfc}\sqrt{\varphi} + 2 \sqrt{\varphi / \pi}\right] \label{pfl}$,
and the contribution of the free and trapped electrons  from the slab R, viz  
$P_\mathrm{R}(\varphi)= \frac{\Theta \Upsilon N_\mathrm{0L}}{2 N_0} \left[ e^{\frac{\varphi}{\Theta}}\mathrm{Erfc}\sqrt{\frac{\varphi}{\Theta}} + 2\sqrt{\frac{\varphi }{\pi \Theta}}+\frac{8}{3} \frac{\varphi^{3/2}}{\sqrt{\pi \Theta^3}}\right]$.
The trapping potential  in Equation (\ref{eq:pot}) was obtained by assuming $\Psi(0)=0$.
From the analogy with particle motion, when the Sagdeev potential (\ref{eq:pot}) is negative the electrostatic potential is driven out of equilibrium and the system supports soliton-like structures. On the other hand, the  conditions of charge neutrality at the boundaries, equivalent to considering $\partial \Psi(\varphi)/\partial \varphi |_{0,\varphi_0}=0$, assure that $\varphi$ grows monotonically from 0 to $\varphi_0$ without oscillating back and forth, while the condition $\Psi(\varphi_0)=\Psi(0)=0$ assures that $\varphi_0$ remains bounded without growing indefinitely.

\begin{figure}
\includegraphics[width=2.5in]{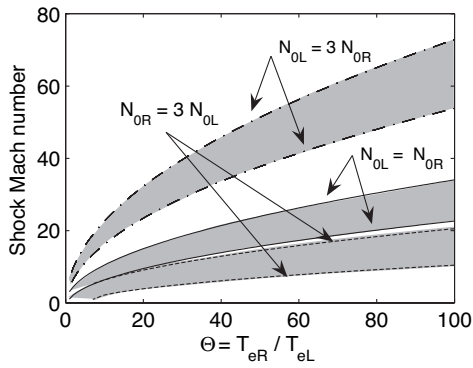}
\caption{\label{fig:maxmac} Maximum Mach number of the shock  as a function of the  electron temperatures ratio  $\Theta =T_\mathrm{eR}/T_\mathrm{eL}$, for three different conditions, namely $N_{0L}=N_{0R}$ (solid line),  $N_{0L}= 3 N_{0R}$ (dash-dot line), and $N_{0L}=N_{0R}$ (dashed line). The shaded areas represent the regions of allowed Mach number.}
\end{figure}

Examining Equation (\ref{eq:pot}), we can thus conclude that the system supports a monotonic double layer solution, for a given Mach number, only if the electron pressure exceeds the ion pressure along the shock, and if both coincide in value and slope at the boundaries $x=x_\mathrm{L}(\varphi=0)$ and $x=x_\mathrm{L}(\varphi=\varphi_0)$.
On the other hand, if the electrostatic potential exceeds the critical value $\varphi_{cr}=M^2/2$, the ion pressure becomes imaginary, and the wave `breaks' \cite{Schamel}.  In order to have a steady state solution, we must then impose  $\Psi(\varphi_{cr})>0$ \cite{Moiseev}. The inequality can be written in terms of ion and electron pressures in the form
$P_e (M^2/2,\Theta,\Upsilon)<P_\mathrm{I}=M^2$, which recovers the same results obtained by previous authors in the limit of $\Theta \rightarrow 1$ and $\Upsilon\rightarrow 1$ \cite{Fors1,Schamel}, and that imposes an upper limit to the Mach number of the shock waves created during the collision of two plasma slabs with temperature ratio $\Theta$ and density ratio $\Upsilon$.  
When $M^2\gg1$ and $M^2\gg\sqrt{\Theta}$, the electron pressure $P_e(\varphi,\Theta,\Upsilon)$ can be properly expanded around $\varphi=\varphi_{cr}$, and the expression for the maximum Mach number can be written as:
\begin{equation}
M_\mathrm{max}=\frac{3(\Upsilon+1)}{\Upsilon}\sqrt{\frac{\pi \Theta}{8}}\label{eq:max}.
\end{equation} 
Equation (\ref{eq:max}) shows that the collision of two plasma slabs can give rise to electrostatic collisionless  shocks with very large Mach number, provided that the two slabs have the appropriate temperature ($\Theta$) and density ($\Upsilon$) ratios. 
As far as we know, no electrostatic shock waves with $M>6$ have been predicted before.  
This difference arises from the the fact that the present model includes the variation of electron pressure not only as a function of the electrostatic potential, but also as function of the temperature jump, $\Theta$,  between the electrons in the downstream and upstream plasma slabs.  When the downstream electrons have a temperature larger than the upstream electrons, the electron pressure in the shock is reduced (few electrons are trapped), the maximum electrostatic potential $\varphi_{cr}$ can stabilize to a larger value defined by $\Theta$, and the maximum Mach number  $M_\mathrm{max}=\sqrt{2 \varphi_{cr}}$ increases accordingly. Figure \ref{fig:maxmac} shows the theoretical prediction of the maximum Mach number as a function of $\Theta$, for different plasma density ratios $\Upsilon$. The solid line represents the collision of two plasma slabs with equal density ($\Upsilon = 1$), recovering the classic  limit $M \sim 3.1$ \cite{Fors1,Fors2} when $\Theta = 1$. The condition for the minimum Mach number is found by imposing that the Sagdeev potential is negative at its minimum. When $\Theta=1$ and $N_\mathrm{0L}=N_\mathrm{0R}$, the minimum Mach number is 1, as in the hydrodynamic limit.

\begin{figure}
\includegraphics[width=2.5in]{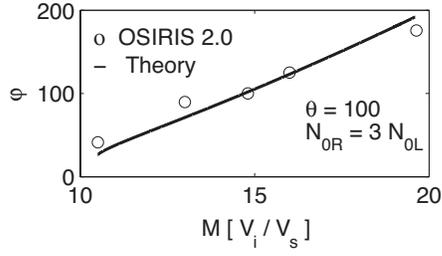}
\caption{\label{fig:pot} Electrostatic potential $\varphi$, normalized to $k_B T_{eL}/ e$, obtained from simulations (circles) and from the theoretical model (solid line), as a function of  the shock wave Mach number $M$.} 
\end{figure}

\begin{figure}
\includegraphics[width=2.5in]{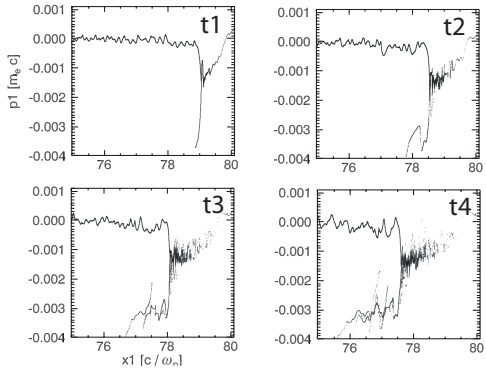}
\caption{Evolution of the ion phase space $p1x1$ at four different times steps $t_1=580.8/\omega_{peL}$, $t_2=871.2/\omega_{peL}$, $t_3=1161.6/\omega_{peL}$ and $t_4=1452.0/\omega_{peL}$. Only the ions from the left slab are shown.  } \label{fig:p1x1}
\end{figure}

In order to check the consistency of the theoretical predictions, we have performed particle-in-cell simulations  using the fully relativistic massively parallel  code OSIRIS 2.0 \cite{Osiris}. The 1D simulations are performed in the reference frame of the slab L,  distance is normalized to $c /\omega_{peL}$, charge to the electron charge $e$, mass to the electron mass $m_e$, and time to $1 /\omega_{peL}$, where $\omega_{peL}=\left(4\pi e^2 N_\mathrm{0L}/m_e \right)^{1/2}$. The box length is $120\,\,c/\omega_{peL}$, with 32768 cells, 50 particles per cell per species (4 species), and the time step is $\omega_{peL} d t=3.63 \times 10^{-3}$. 
The simulations start at $\omega_{peL}\,t=0$, with the slab L occupying the region $x\,\omega_{peL}/c =[0,80]$,  and the slab R occupying the region $x\, \omega_{peL}/c =[80,120]$. In the simulations, the shock is driven by the slab R, both moving to the left; the shock and driver velocities are calculated in the frame of the slab L. The simulations cover a wide range of parameters, with the driver velocity varying between Mach 2 and Mach 40.  
The electrons in the  R and L slabs have temperatures $T_\mathrm{eR} = 1$ keV and $T_\mathrm{eL} = 10$ eV, respectively, the temperature ratio is $\Theta = 100$, and the density ratio is $\Upsilon=3$. In such conditions of temperatures and densities, depending on the  velocity of the driver, the theory predicts the formation of shock waves with Mach number ranging between 10 and 20 (cf. Figure \ref{fig:maxmac}).  Figure \ref{fig:pot} shows the comparison between the theoretical and the numerical electrostatic potential $\varphi$, as a function of the  Mach numbers of the shocks observed in the simulations; such shocks showed velocities ranging between 10 and 20 times the ion sound speed, in excellent agreement with the theory. It should be noted that, while no shocks are observed for drivers (slab R) moving with velocity above Mach 20, very high Mach number shocks ($M \sim 10 - 20 $) are created by drivers moving with velocity between Mach 2 and  Mach 20. In the simulations, the value of the electrostatic potential has been calculated for well developed shock structures, i.e. normalized times much larger than $\omega_{peL}\,t=1000$.  
Figure \ref{fig:p1x1} shows the typical phase space $p1x1$ of the ions from the L slab, at four different time steps. The slab R moves towards the left with $M= 15$, and drives an electrostatic shock moving at $M=16$. The electrostatic potential predicted by the theory for the conditions in our simulation, $\varphi_T=123$, is in very good agreement with the electrostatic potential obtained in the simulations, $\varphi_O=125 \pm 5$. As the shock structure propagates, the ions are picked up and accelerated: a small fraction  is reflected by the electrostatic potential, while most of the ions end up with the same speed of the shock wave. We have also performed simulations of scenarios with two identical slabs: in these scenarios, when colliding at $M>6$, no shock formation was observed, thus confirming that high Mach number shocks are supported only when the colliding slabs have the appropriate temperature and density ratios. 

In conclusion, we have shown theoretically and numerically that very high Mach number shock waves are formed during the collision of plasma slabs. The simulations confirmed that such shock waves, which travel with Mach numbers well above previous predictions \cite{Sagdeev,Schamel,Fors1,Fors2, MongJoyce}, arise naturally during collision of plasma slabs with different electron temperatures, and driver velocity between $M=2$ and $M=M_{max}$. Such situations could readily arise in astrophysics and in the interaction of high intensity lasers with plasmas \cite{Silva}. We have also shown that the maximum Mach  number grows  with the electron temperature ratio as $\sim \Theta^{1/2}$, and with the electron density ratio as $\sim (1+\Upsilon)/\Upsilon$.
Further theoretical and numerical analysis will extend the theory to extremely large Mach number shocks for which $M\gtrsim V_{TeL}/V_s$, and include the influence of relativistic effects for shock velocities comparable with the speed of light.
  
\begin{acknowledgments}
This work was partially supported by FCT (Portugal) through grants POCTI/FP/FAT/501900 and POCTI/FIS/55095. G.S. is supported by FCT (Portugal) though the scholarship SFRH/BPD/17858/2004. G.Sorasio would like to thank Prof. H. Schamel for suggestions, and the Abdus Salam ICTP, Trieste for the kind hospitality. LOS acknowledges useful discussions with Prof. Warren Mori and Prof. Chuang Ren. The simulations were performed in the expp cluster at IST.
\end{acknowledgments}

\newpage

\end{document}